\documentstyle[aps,pre,graphicx]{revtex} 
\begin{document}
\title{Many-particle simulation of the evacuation process from a room without visibility} 

\author{Motonari Isobe$^{a}$, Dirk Helbing$^{b}$, Takashi Nagatani$^{a}$}

\address{$^{a}$Department of Mechanical Engineering, Division of Thermal Science, 
Shizuoka University, Hamamatsu 432-8561, Japan\\
$^{b}$ Institute for Economics and Traffic, Dresden University of Technology, 
D-01062 Dresden, Germany} 
\maketitle

\begin{abstract}
We study the evacuation process from a smoky room by means of experiments 
and simulations. People in a dark or smoky room are mimicked by ``blind'' students 
wearing eye masks. The evacuation of the disoriented students from the 
room is observed by video cameras, and the escape time of each student is 
measured. We find that the disoriented students exhibit a distinctly different 
behavior compared with a situation in which people can see and orient themselves.
Our experimental results are reproduced by an extended lattice gas model taking into account
the empirically observed behavior. Our particular focus is on the mean value and distribution
of escape times. For a large number of people in the room, the escape time distribution
is wide because of jamming. Surprisingly, adding more exits does not 
improve the situation in the expected way, since most people use the exit that is discovered first, 
which may be viewed as ``herding effect'' based on accoustic interactions. 
Moreover,  the average escape time becomes minimal for
a certain finite number of people in the dark or smoky room. These non-linear effects have practical
implications for emergency evacuation and the planning of safer buildings.
\end{abstract}

\section{Introduction} 

During the last decade, many observed self-organization phenomena in traffic 
flows \cite{Ref1,Ref2,Ref3,Ref4,Ref5} and pedestrian streams 
\cite{Ref6,Ref7,Ref8,Ref9} have been successfully reproduced 
with physical methods. This has not only stimulated research in granular, 
biological, and colloid physics \cite{Ref10,Ref11,Ref12}. It has also encouraged physicists 
to study evacuation processes \cite{Ref13,Ref14,Ref15}, since it has been shown that many 
aspects of crowd stampedes can be understood by driven many-particle models 
\cite{Ref16,Ref17}. The empirical observations have many common features with driven 
granular media.

Evacuation processes have been studied by the use of various simulation 
models. The typical models of pedestrian motion are based on molecular 
dynamics methods \cite{Ref6,Ref16,Ref17}, lattice gas models \cite{Ref7,Ref18,Ref19,Ref20,Ref21,Ref22,Ref23,Ref24}, or cellular 
automata \cite{Ref9,Ref14,Ref15,Ref25}. In simulations of evacuation processes, it 
has been mostly assumed that visibility is good. However, this is often not the case 
during fire emergencies, as smoke or a failure of the electrical power supply reduce the 
orientation significantly. Because of the toxic effects of smoke, fast evacuation is particularly
important, but little is known about the behavior of people under conditions of bad or no visibility.

In this study, we will therefore focus on the investigation of the 
evacuation process from a smoky room without visibility. 
In favour of video recordings, we mimic this situation by requiring our test persons to 
wear eye masks. This allows us to study not only qualitative features, 
but also quantitative outcomes, which is relevant for reliable predictions of evacuation times
and the planning of safer buildings or pedestrian facilities.
From our video recordings, we have evaluated the characteristic behavior and
the escape times in well-controlled experiments (see Sec.~\ref{Sec2}). 
In Section~3, these experiments are reproduced with an extended lattice 
gas model of pedestrian flows with model parameters calibrated to the data. 
Despite of the simplicity of the model, it can successfully reproduce 
the characteristic escape behavior under conditions of no visibility
and the empirical escape times in a semi-quantitative way. Moreover, we identify
two interesting effects: First, the average escape time becomes minimal for
a specific finite number of people, who are initially in the room. Second, adding
more exits does not increase the efficiency of evacuation in the expected way. 

\section{Experiment} \label{Sec2}

We have experimentally studied the evacuation of blind students from an 
empty classroom, which is schematically illustrated in Figure~{1}. Each 
student wore an eye mask. The exact width of the classroom was $W=4.2$~m and 
its length $L = 5.5$~m. There were no obstacles in the classroom, i.e. desks and chairs
were moved aside (to form the boundary). Moreover, the room had one exit of width 0.5 m. Two video 
cameras 1 and 2 were located within and in front of the classroom. A 
cameraman was able to observe all the students by video camera 1. The other 
cameraman could observe the students who escaped through the exit by video 
camera 2. We investigated two cases: 
\begin{itemize}
\item[(a)] the escape of only one disoriented 
student with an eye mask and 
\item[(b)] the evacuation of 10 disoriented students. 
\end{itemize}
Correspondingly, at 
time $t=0$, there were either (a) one student or (b) 10 students in the classroom, and each student 
was standing at a random place within a central area of the room. Before the experiment, all students were forced by 
the cameraman to turn around themselves. In the result, they lost their 
directional orientation. All students moved to seek for the exit as soon as 
a cameraman shouted a word of command. The evacuation process was then 
recorded by the two video cameras.

We have first studied the case (a) of single disoriented students with eye masks. By careful analysis 
of the video recordings, we have determined the trajectory and escape time 
of each student. The individual escape time was defined as the time elapsed 
between the shouting of the command and the moment when the respective 
student left the room through the exit. Figure~{2} shows four typical 
trajectories of a single student until he successfully escaped from the room. His 
face direction at the initial position is indicated by an arrow. At first, the student 
turned slightly, and then he moved slowly towards one of the walls of the room. As 
soon as he touched the boundary, he followed it, after he had chosen the right-hand or left-hand direction 
at random, because of the loss of directional memory. 
As soon as he found the exit, he left the room. Note that the following of the wall differs significantly
from the behavior assumed in the evacuation simulations of Ref.~\cite{Ref16}, where individuals
``hitting'' a wall where ``reflected''. It is, therefore, interesting and important to check, whether
the conclusions for the evacuation of a group of people are to be revised. It will turn out that
our new simulations are an independent, experimentally supported confirmation of the previous
findings, which were obtained for a hypothetical pedestrian behavior under conditions of smoke. 

In Figure~{2}, the escape time obtained from the experiment is shown below each trajectory. 
One can see that the disoriented students sometimes take the shorter way and at other times
the longer way. Thus, the escape time depends highly on the randomly chosen direction. We repeated 
the experiment 10 times with 10 different students. The mean escape time of $t_e = 
31.5\sec $ was obtained by averaging over all experiments, but the variance was large, as expected. 

Next, we have studied the case (b) of 10 disoriented students with eye masks. Initially, the 10 students 
were standing at random places close to the center of the room. By careful analysis 
of our video recordings, we have determined the trajectories and escape 
times of all 10 students. Figure~{3} shows a photo of the evacuation of students 
from the room at time $t=16$~sec. The students move along the wall. Figure~{4} shows the 
time evolution of the evacuation process for 10 students. The patterns 
(a)-(d) were obtained at times $t=0$~sec, 5~sec, 10~sec, and 15~sec. Numbered circles 
represent the 10 disoriented students, whose face directions are indicated by 
arrows. The students turned slightly at first, and moved slowly towards one of the 
walls of the room. As soon as they touched a wall of the room, they followed it in one of
the two possible directions. However, as soon as one or two students managed to leave the room, the 
remaining students noticed the location of the exit accoustically. Then, the remaining 
students managed to find the exit much faster. When a student met 
another one, they went together along the wall into the same direction. They 
left the room as soon as they reached the exit.

We determined the escape times of all students by careful analysis of our 
video recordings of 10 repetitions of the experiment. The average escape time of the ten 
students was 22.1 sec. However, the escape time of the fastest student was 
9.8 sec, while the value of the slowest student was 34.3 sec. Thus, the 
distribution of escape times was again rather wide. Notice that the average escape 
time of 22.1 sec for 10 students is significantly lower than the 31.5 sec 
for a single student! This lower value of escape time is due to the reason 
that, when one or two students managed to leave the room successfully, the 
remaining students noticed the location of the exit and inverted their direction, if appropriate.

For comparison, we performed an escape experiment for 10 students
without eye masks corresponding to a room with normal visibility. Figure~{5} shows 
the time evolution of the evacuation process for all 10 students. The patterns 
(a)-(d) were obtained at times $t=0$~sec, 1~sec, 3~sec, and 5~sec. Their face directions 
are indicated by arrows. The students turned instantly towards the exit and 
then moved into its direction very fast. When the arrival rate of students 
exceeded the capacity of the exit, students were queueing, i.e. a crowd of 
students was forming in front of the exit due to jamming. The average escape 
time of the 10 students was 6.54 sec. This value is significantly lower than 
the 22.1 sec for 10 disoriented students with eye masks. 

Therefore, the evacuation process from a room with no visibility is qualitatively and 
quantitatively different from the evacuation under normal conditions. For this reason, it 
is necessary and important to develop a simulation model of the evacuation of people,
which allows to assess the efficiency of escape and the safety of buildings
under different conditions, in particular under conditions of no visibility.

\section{Many-Particle Simulation} \label{Sec3}

In the following, we will describe a simple model which allows to reproduce our 
experimental findings for the evacuation of a smoky room in a 
semi-quantitative way. Here, we will simulate the pedestrian flow by the use of a 
lattice gas model, but it would be also possible to use the social force model of
pedestrian behavior \cite{Ref6,Ref16,Ref17}. We have implemented the following characteristics of
disoriented people: 
\begin{itemize}
\item[(1)] Each person turns slightly at first and moves 
towards one of the the walls. 
\item[(2)] He chooses the right-hand or left-hand direction at random as soon 
as he reaches a wall. 
\item[(3)] Afterwards, he moves along the wall. 
\item[(4)] When one 
or two students have managed to leave the room (after the ``exploration phase''), 
the remaining students turn into the direction of the exit as well. 
\end{itemize}
Each disoriented student is represented by a walker 
on a square-diagonal lattice with $L$x$W$ sites reflecting the classroom. 
We choose the lattice spacing as 0.4 m, since the 
typical space occupied by a pedestrian in a dense crowd is about 0.4 m x 0.4 
m. We therefore use $L=14$ and $W=11$. The classroom is connected to the outer 
space through a single exit represented by one site. Figure~{1} shows a 
schematic illustration. An empty circle represents a student in the 
classroom. In reasonable agreement with the empirical observations, we assume that each walker 
performs a biased-random walk on the square-diagonal lattice until he 
reaches the boundaries of the room \cite{onedoor}. Initially, each walker chooses randomly 
one of the eight directions on the square-diagonal lattice. The walker is 
then biased with respect to this direction, which represents the desired 
walking direction of the student. The biased random walker is allowed to move 
not only to the nearest-neighbor sites, but also to the next-nearest 
neighbor sites (in the diagonal directions). 

Figure~{6} illustrates two of all possible configurations of a biased random 
walker on the square-diagonal lattice, where bias is assumed in the upward 
direction. Configuration (a) shows the situation of being able to move to all 
the nearest-neighbor and next-nearest-neighbor sites, when these are not occupied 
by other walkers. The arrows indicate the possible directions in which the 
walker can move. The transition probabilities of the walker into the 8 
directions are given by

\[
p_{t,y} = D / 3 + (1 - D) / 8 = p_{t,y}^{\rm (a)} 
\]

\[
p_{t,x,y} = p_{t, - x,y} = D / 6 + (1 - D) / 8 = p_{t,x,y}^{\rm (a)} 
\]

\[
p_{t,x} = p_{t, - x} = D / 9 + (1 - D) / 8 = p_{t,x}^{\rm (a)} 
\]

\begin{equation}
\label{eq1}
p_{t, - y} = p_{t,x, - y} = p_{t, - x, - y} = D / 27 + (1 - D) / 8 = p_{t, -y}^{\rm (a)}  \, ,
\end{equation}
where $D$ is the parameter representing the bias, $p_{t,y}$ is the transition probability 
in $y$-direction, and the superscript (a) refers to configuration (a). 
Specifically, $p_{t,y}^{\rm (a)}$ indicates the transition probability into the 
$y$-direction in configuration (a), $p_{t,x,y}^{\rm (a)}$ the transition probability into
the $x$- and $y$-direction, i.e. into diagonal direction, etc. In the following simulation, 
we set $D$=0.99. The transition probability is then $p_{t,y}^{\rm (a)}  = 0.33$. 
This  value agrees with the average speed 0.33 m/s obtained from the experiment. 
Therefore, we identify one time step in our simulations with one second. 

Figure~{6}(b) shows the configuration in which another walker occupies one of the 
nearest-neighbor sites. The location of the other walker is indicated by a cross. 
Then, the transition probabilities into the remaining 7 directions are given by

\[
p_{t,x,y} = p_{t, - x,y} = p_{t,x,y}^{\rm (a)}  + \frac{p_{t,x,y}^{\rm (a)} }{(2p_{t,x,y}^{\rm (a)}  + 
2p_{t,x}^{\rm (a)}  + 3p_{t, - y}^{\rm (a)}  )}\times p_{t,y}^{\rm (a)}  / 7
\]

\[
p_{t,x} = p_{t, - x} = p_{t,x}^{\rm (a)} + \frac{p_{t,x}^{\rm (a)} }{(2p_{t,x,y}^{\rm (a)} + 
2p_{t,x}^{\rm (a)} + 3p_{t, - y}^{\rm (a)} )}\times p_{t,y}^{\rm (a)} / 7
\]

\begin{equation}
\label{eq2}
p_{t, - y} = p_{t,x, - y} = p_{t, - x, - y} = p_{t, - y}^{\rm (a)} + \frac{p_{t, - 
y}^{\rm (a)} }{(2p_{t,x,y}^{\rm (a)}+ 2p_{t,x}^{\rm (a)} + 3p_{t, - y}^{\rm (a)} )}\times p_{t,y}^{\rm (a)} / 7.
\end{equation}

For the other possible configurations, the transition probabilities of walkers 
are specified analogously. Therefore, the explicit 
expressions are omitted to safe manuscript space.

In our simulations, we assume that initially (at time $t = 0$), all disoriented 
students stand at some location in the classroom without any directional 
memory. In the next time step ($t = 1)$, each student starts moving in order 
to escape from the room. Until he reaches the boundaries (a wall), he performs a biased 
random walk according to the model sketched above. For each random 
walker, we assume a constant bias into the desired direction defined by the first step. All walkers are updated 
once every time step in a random sequential way as, in reality, the students move 
asynchronously (in contrast to the synchronized movement of soldiers in a military corps). 
When a walker reaches the wall, he chooses the direction to the left or to the right
randomly with probability 1/2. After this choice,  he moves along the wall. 
Furthermore, when one walker manages to leave the room, the remaining walkers adopt their desired 
direction to the direction of the exit. When a walker reaches 
the exit, he is removed from the simulation. Excluded volume effects are taken 
into account by preventing multiple occupation of the same site, i.e. each 
site contains only one individual walker or it is empty. 

Our numerical results are displayed in 
Figs.~{7} to {13}. Figure~{7} 
shows representative trajectories when there is only one walker. The 
related escape times for trajectories (a) to (d) were (a) $t_e = 31\sec $, 
(b) $t_e = 32\sec $, (c) $t_e = 41\sec $, and (d) $t_e = 23\sec $. These 
values are indicated below the trajectories. For illustrative reasons,
from the many simulated trajectories, we have selected the examples (a)-(d) which looked similar to 
the ones displayed in Fig.~{2}, which were obtained experimentally. The mean 
value of 30.1 sec for the simulated escape time was obtained by averaging 
over 100000 samples, as compared to an average escape time of 31.5 sec in 
our experiments. Figure~{8} shows the probability density of escape times for 
one student, obtained from 100000 simulation runs. The probability density 
exhibits a rather wide distribution, but the mean escape time agrees well 
with the one obtained from the experiment.

Figure~{9} shows the time evolution of the evacuation process of 10 persons 
according to our simulations. The representative patterns (a)-(d) were obtained 
at times $t=0$~sec, 
5~sec, 10~sec, and 15~sec. Full circles represent 10 walkers, whose face 
directions are indicated by arrows. 
The behavior of the 10 simulated walkers was qualitatively the same as the one observed in 
our experiments (see Fig.~{4}). The mean value of the escape time was reduced 
as in our experiments, but with 27.1~sec, the simulation value was 
a pessimistic estimate of the experimental value. 

Similarly, Figure~{10} shows the simulated time evolution of the evacuation 
process for 20 persons. The patterns (a)-(d) were obtained at times $t=0$~sec, 5~sec, 
10~sec, and 20~sec. The behaviors of the 20 simulated walkers was similar to those 
of 10 persons. However, jamming appeared near the exit at $t=20$~sec,
because the arrival rate of walkers exceeds the capacity of the exit.

We have determined the probability density distribution of escape times of a 
finite number of walkers from 100000 simulation runs. Figures~{11}(a)--(d) 
show, respectively, the plots of the probability density of escape times for 
5, 10, 15, and 20 persons. In each figure, the probability density 
distributions are shown for the 1st, 5th, 10th, 15th, and 
20th person. Each probability density distribution has approximately a 
Gaussian shape. With an increasing number of walkers, the probability 
density distribution of escape times for the $n$th person becomes 
narrower, i.e. the variance decreases. Figure~{12} shows the probability 
density distributions of the overall escape times for all persons in the cases of 1, 5, 
10, 15, and 20 walkers. Up to 10 walkers, the probability density 
distribution becomes slightly narrower with an increasing number of walkers, 
but the distribution becomes rather wide when the number of walkers exceeds 
10. For less than 10 walkers, the  efficiency of the escape is enhanced by the presence of 
other persons who may discover the exit, but the escape is obstructed by the 
other persons for more than 10 walkers, which is due to jamming near the 
exit. 

Let us now study the simulated escape time of the first student who successfully escapes from a 
room with no visibility. Figure~{13} shows the plot of the first walker's mean escape time 
as a function the overall number of walkers initially present in the room. With 
an increasing number of walkers, the escape time of the first escaped walker decreases. 
This is, because the chances to discover the exit increase with the presence of more people. 

\section{Simulations with Two Doors} \label{Sec4}

We will now study the effect of two exits on the evacuation process from a room with no
visibility (e.g. a room with heavy smoke)
by means of simulations. The second exit is assumed to be located on the opposite side of
the room. 

The simulations basically agree with the ones for one exit, but we will consider two different cases: 
In scenario A, as soon as one of the walkers has found an exit, the other walkers are assumed to 
recognize the location of this exit accoustically and to move into its direction, as in the scenario
with one exit. However, in scenario B we assume that walkers do not recognize or ignore the discovery of an
exit by other walkers. In scenario A, the second door is practically unused, as the walkers turn towards
the exit discovered first. Consequently, an additional door does not double the flow
of escaped persons and does not reduce the average escape time by a factor of two, as planners would usually
assume. It mainly reduces the exploration time until a door is discovered. In contrast, in scenario B 
walkers use the two alternative exits approximately with the frequency that planners usually assume. 
Figure~{14} shows the average
evacuation time for a room with one and two doors, respectively. Circles, squares, and triangles indicate, 
respectively, the simulation results for the situation with one exit, for scenario A and scenario B. 
Figure~{15} displays the overall evacuation time as a function of the initial number of walkers. 
When the initial number of persons is increased, the difference in the escape times between scenarios 
A and B becomes large.
This result urgently calls for an accoustic guidance of people towards the exit in evacuation situations,
in order to use the evacuation capacities of all available exits, and avoid unnecessary jamming. In
situations with normal visibility, there is a tendency to have a load balancing between
alternative exits \cite{twodoor}. 

\section{Summary} \label{Sec5}

Focussing on the individual escape times, we have presented 
experimental results on the evacuation of disoriented students 
from a classroom with no visibility. The behavior of 
the disoriented students was very characteristic. They moved slowly towards 
one of the walls of the room and then along the wall. As soon as one of the persons managed
to leave the room, the other ones recognized the location of the exit accoustically and moved
into this direction. This reduced the average escape time in the case of one door,
but it produced unnecessary jamming in the case of multiple doors. Therefore, accoustic
guidance towards the doors would significantly 
increase the efficiency of usage of alternative doors, decrease the average escape
times, and increase the chances of survival in emergency situations. 

Despite of the stochastic nature of 
pedestrian flows, the empirical observations could be semi-quantitatively 
reproduced by an extended lattice gas model, i.e. a stochastic many-particle 
approach. In particular, we could successfully reproduce the empirically observed behavior of 
persons in a room  without visibility: For example, the average escape 
time per person becomes minimal for a certain number of persons (see Fig.~{14}), as 
the chances to find the exit increase with the number of persons (see Fig.~{13}), but 
obstructions due to jamming at the exit dominate for a higher number of persons (see Fig.~{12}). 
As a consequence, the model could be used to identify not only the expected 
escape times as a function of the number of people in a dark or smoky room, 
but it could also help to identify the probability distribution of escape 
times. This is of practical importance for the assessment of the safety of buildings in emergency
situations.

\clearpage

\section*{Figure Captions}

\bigskip

FIG. 1\textbf{.} Schematic illustration of the classroom of width $W=4.2$ m and 
length $L=5.5$ m, in which desks, chairs, and other obstacles have been moved aside. There is only 
one exit in the front of the classroom, the width of which is 0.5 m. Two 
video cameras 1 and 2 were located within and in front of the classroom. In 
our simulations, a square-diagonal lattice of $W=11$ sites and $L=14$ sites is 
used.

\bigskip

FIG. 2. Four typical trajectories of a student until he successfully escapes 
from the room. His face direction at the initial position is indicated by an 
arrow. Below each trajectory, the escape time obtained from the experiment 
is indicated.

\bigskip

FIG. 3. Photo of the evacuation of students from the room at $t=16$ sec.

\bigskip

FIG. 4. Time evolution of the evacuation process of 10 students without 
visibility of the exit. The patterns (a)-(d) were obtained at $t=0$~sec, 5~sec, 
10~sec, and 15~sec. Numbered circles represent 10 disoriented students with eye masks, whose face 
directions are indicated by arrows.

\bigskip

FIG. 5. Time evolution of the evacuation process for 10 students in a room 
with normal visibility. The patterns (a)-(d) were obtained at $t=0$~sec, 1~sec, 3~sec, 
and 5~sec. Their face directions are indicated by arrows.

\bigskip

FIG. 6. Two of all possible configurations of a biased-random walker on the 
square-diagonal lattice, where bias is applied to the upward direction. 
Configuration (a) shows the situation of being able to move to all the 
nearest-neighbor and next-nearest-neighbor sites. The arrows indicate the 
possible directions in which the walker can move. Configuration (b) shows the 
situation in which another walker occupies one of the nearest-neighbor sites, which is
indicated by a cross. 

\bigskip

FIG. 7. Representative trajectories obtained from the simulation when there 
is only one walker. The escape times obtained are shown below the 
trajectories.

\bigskip

FIG. 8. Plot of the probability density of the simulated escape time for a single walker.

\bigskip

FIG. 9. Simulated time evolution of the evacuation process of 10 persons. 
The patterns (a)-(d) were obtained at $t=0$~sec, 5~sec, 10~sec, and 15~sec. Full 
circles represent 10 students, whose face directions are indicated by 
arrows.

\bigskip

FIG. 10. Simulated time evolution of the evacuation of 20 persons. The 
patterns (a)-(d) were obtained at $t=0$~sec, 5~sec, 10~sec, and 20~sec.

\bigskip

FIG. 11. Plots of the probability density of the simulated escape times for (a) 
5, (b) 10, (c) 15, and (d) 20 persons. In each figure, the probability 
density distributions are shown for the 1st, 5th, 10th, 
15th, 20th, and all persons.

\bigskip

FIG. 12. Probability density distributions of the simulated escape times for all 
persons in the case of 1, 5, 10, 15, and 20 walkers.

\bigskip

FIG. 13. Plot of the mean escape time of the first escaped walker as a function of the overall 
number of walkers who are initially present in the room.

\bigskip

FIG. 14. Plot of the average escape time as a function of the initial number of walkers. Circles represent
the results for one single exit. In the 2-exit simulation, the average escape time is only reduced a little,
as walkers orient towards the exit which has been discovered first, which produces jamming at one door 
(scenario A, see the squares). The
expected significant reduction in the average escape time is only found, if walkers do not react to the
discovery of a door by other walkers (scenario B, see the triangles).

\bigskip

FIG. 15. Plot of the overall evacuation time of all walkers as a function of the 
initial number of persons. A second exit reduces the overall evacuation time with respect to the
situation with one exit (circles), but it does not reduce it by a factor of 2. If people orient towards the first
discovered door (scenario A), the overall escape time is mainly reduced by the earlier discovery of
an exit (squares). However, the overall escape time is normally much higher 
than in the hypothetical scenario B, for which
we assume that the discovery of an exit by other people is not recognized or ignored, 
and both exits are equally used (triangles). 
\clearpage

\begin{figure}[htbp]
 \begin{center}
  \includegraphics[width=12cm]{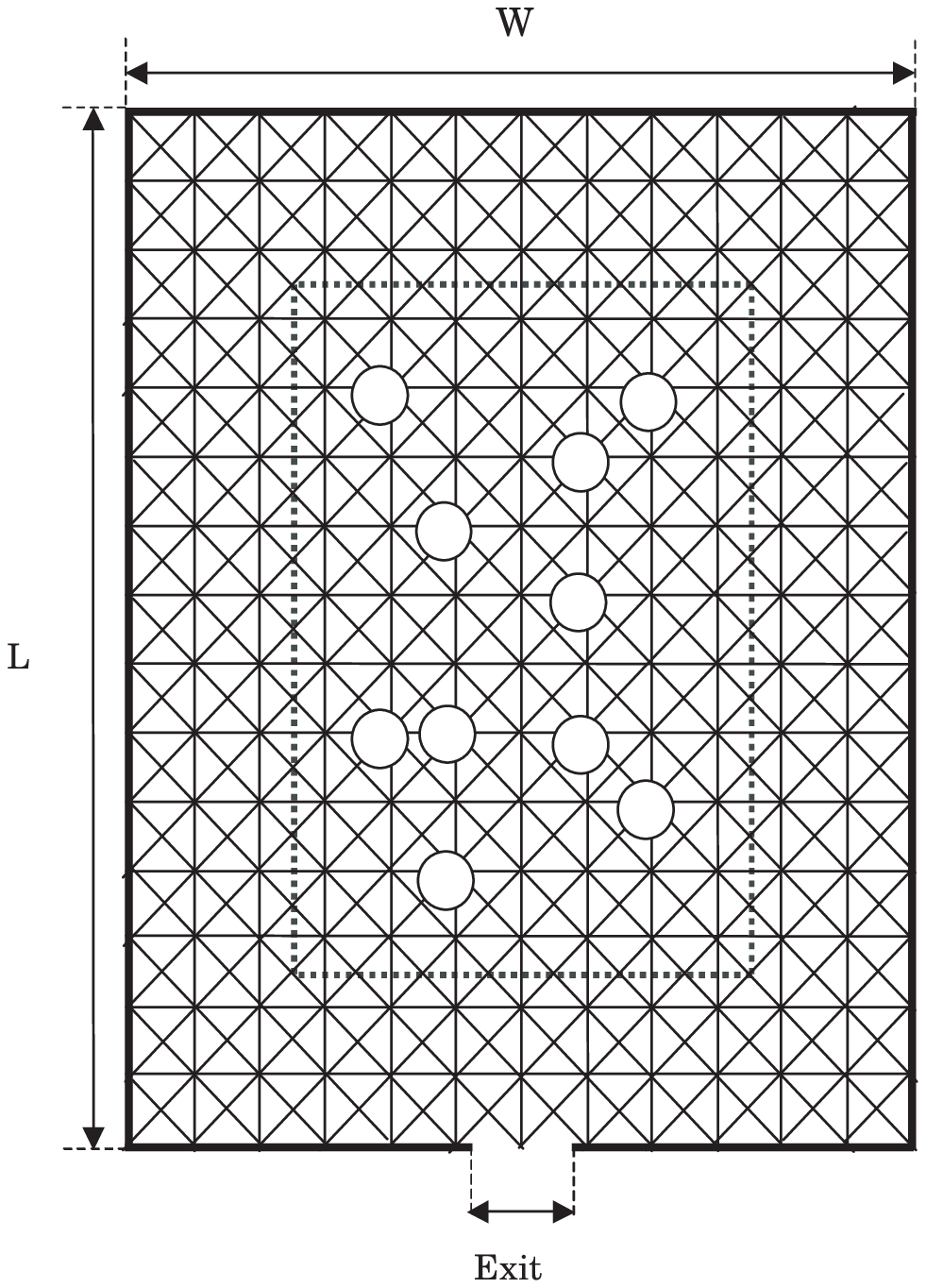}
 \end{center}
\caption[]{\label{Fig1}}
\end{figure}

\begin{figure}[htbp]
    \begin{center}
        \includegraphics[width=12cm]{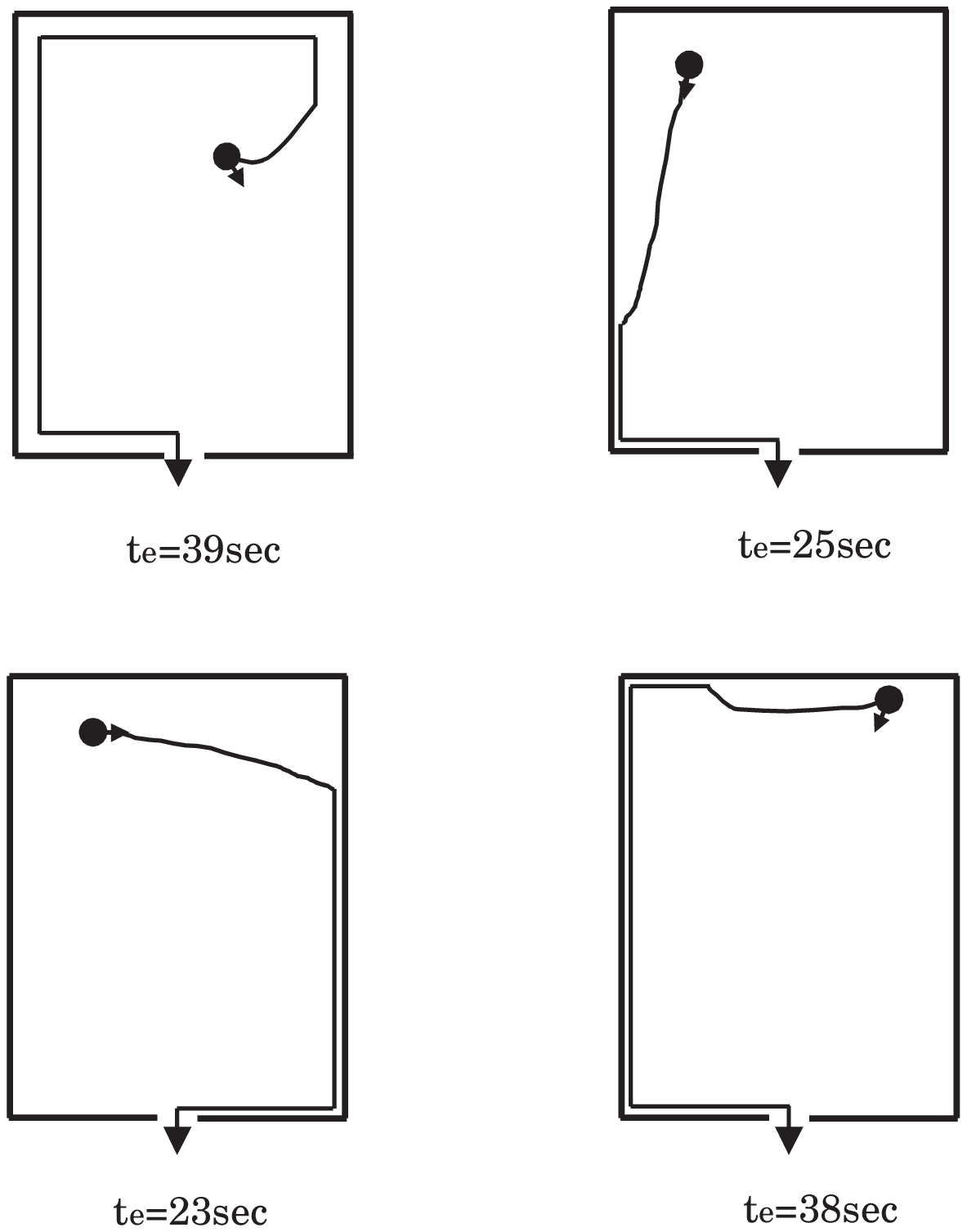}
    \end{center}
\caption[]{\label{Fig2}}
\end{figure}

\begin{figure}[htbp]
    \begin{center}
    \end{center}
\caption[]{\label{Fig3}}
\end{figure}

\begin{figure}[htbp]
    \begin{center}
        \includegraphics[width=12cm]{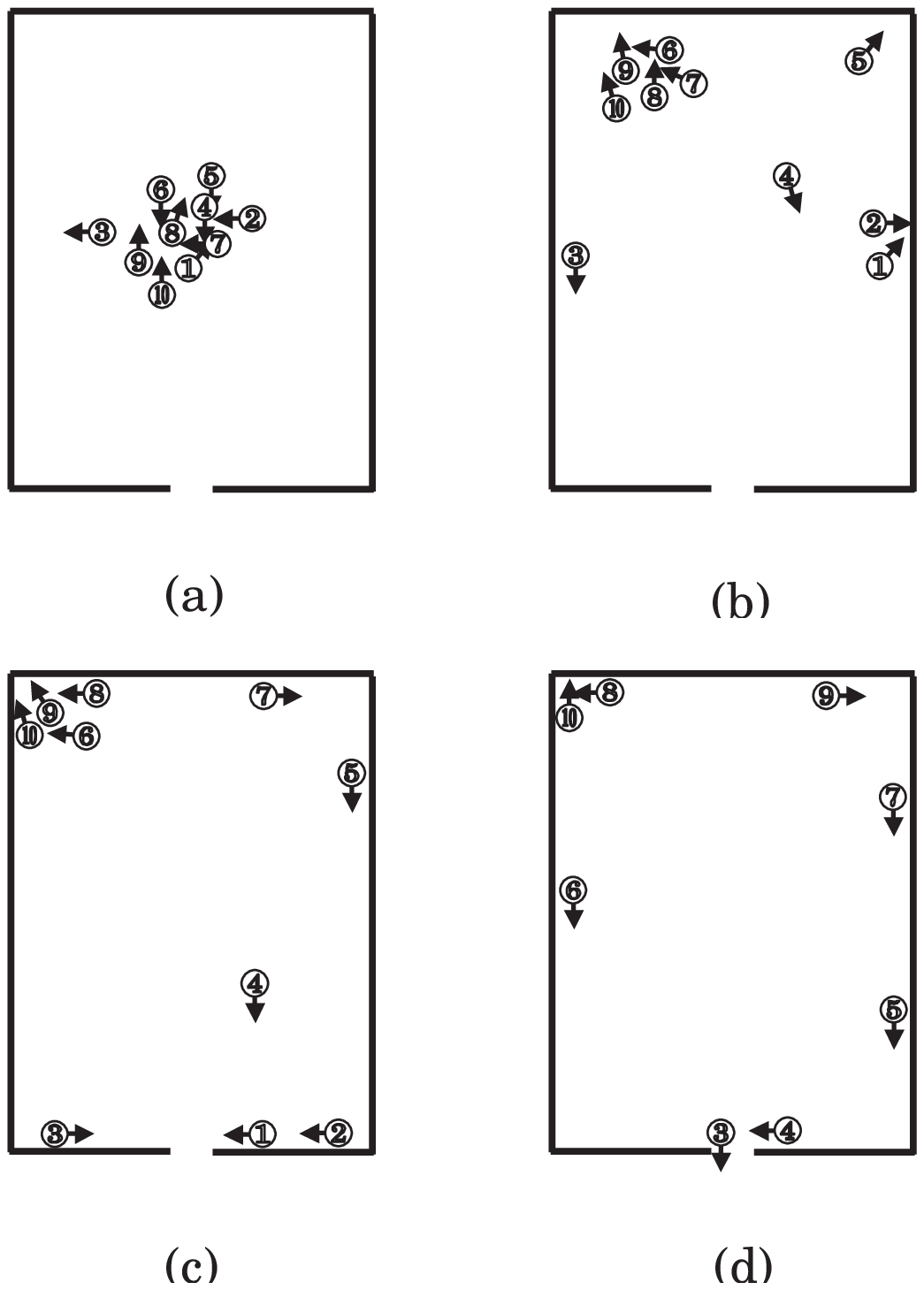}
    \end{center}
\caption[]{\label{Fig4}}
\end{figure}

\begin{figure}[htbp]
    \begin{center}
        \includegraphics[width=12cm]{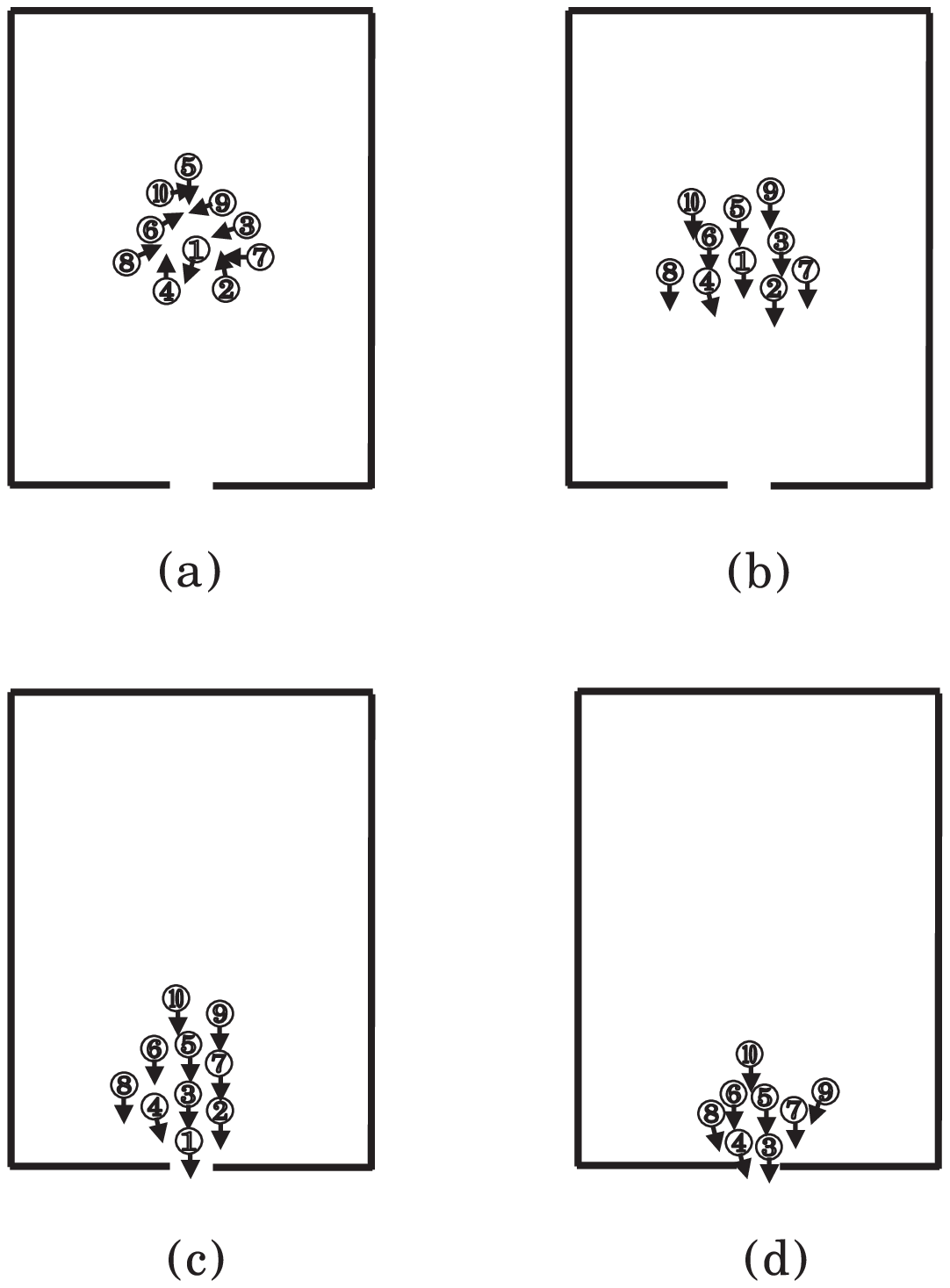}
    \end{center}
\caption[]{\label{Fig5}}
\end{figure}

\clearpage

\begin{figure}[htbp]
    \begin{center}
        \includegraphics[width=12cm]{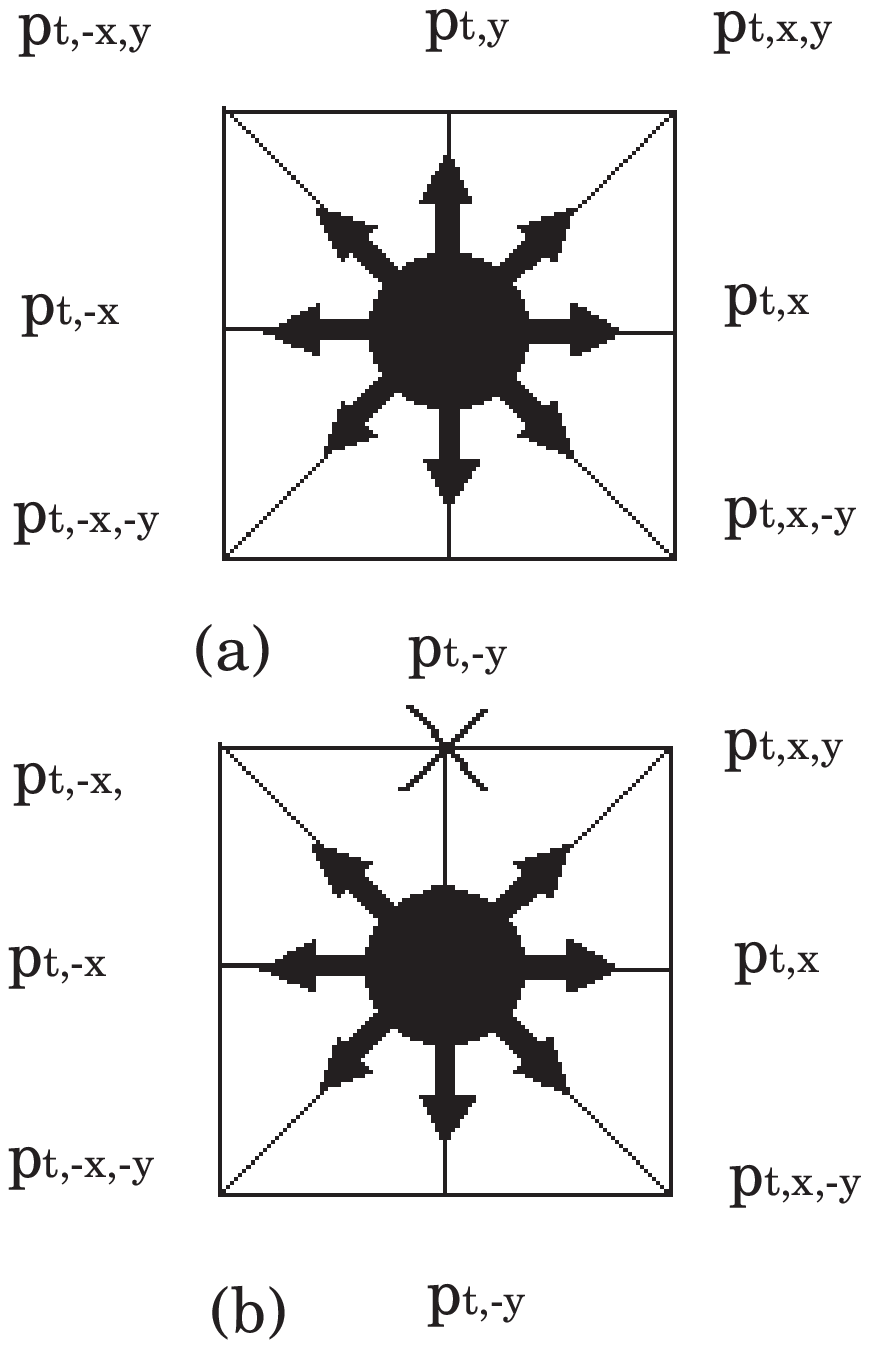}
    \end{center}
\caption[]{\label{Fig6}}
\end{figure}

\begin{figure}[htbp]
    \begin{center}
        \includegraphics[width=12cm]{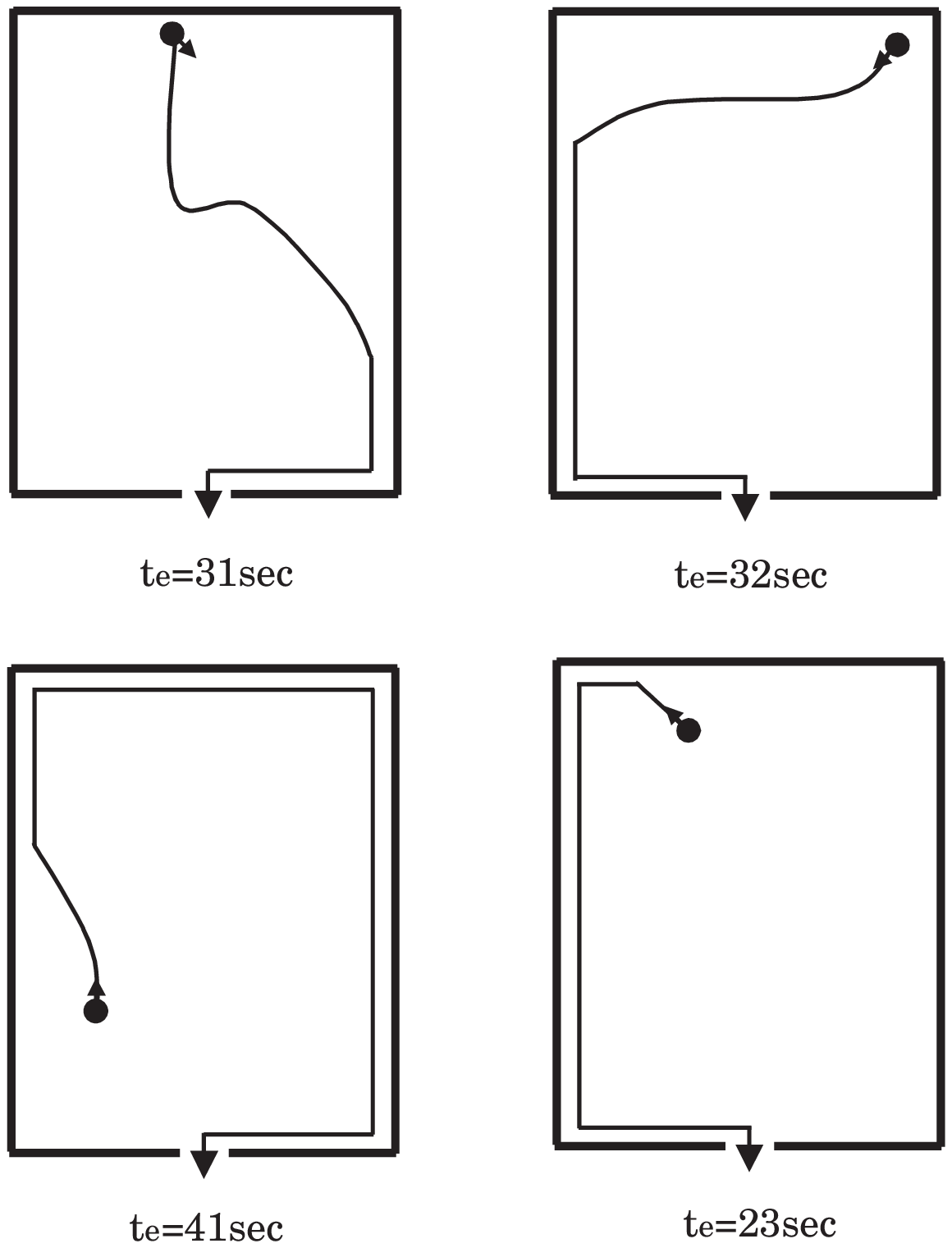}
    \end{center}
\caption[]{\label{Fig7}}
\end{figure}

\clearpage

\begin{figure}[htbp]
    \begin{center}
        \includegraphics[width=12cm]{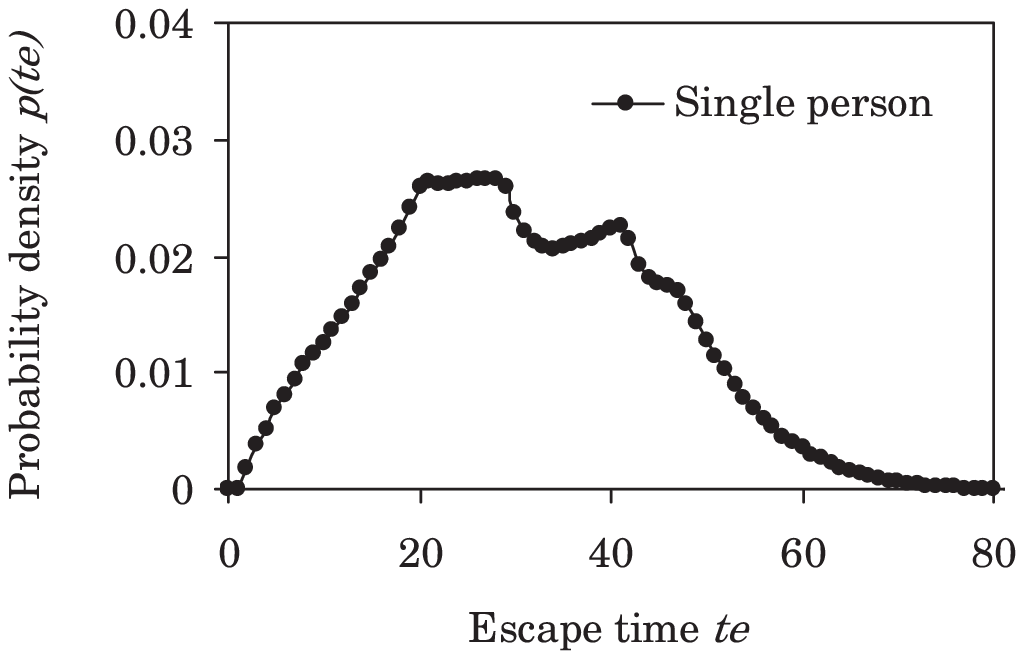}
    \end{center}
\caption[]{\label{Fig8}}
\end{figure}

\begin{figure}[htbp]
    \begin{center}
        \includegraphics[width=12cm]{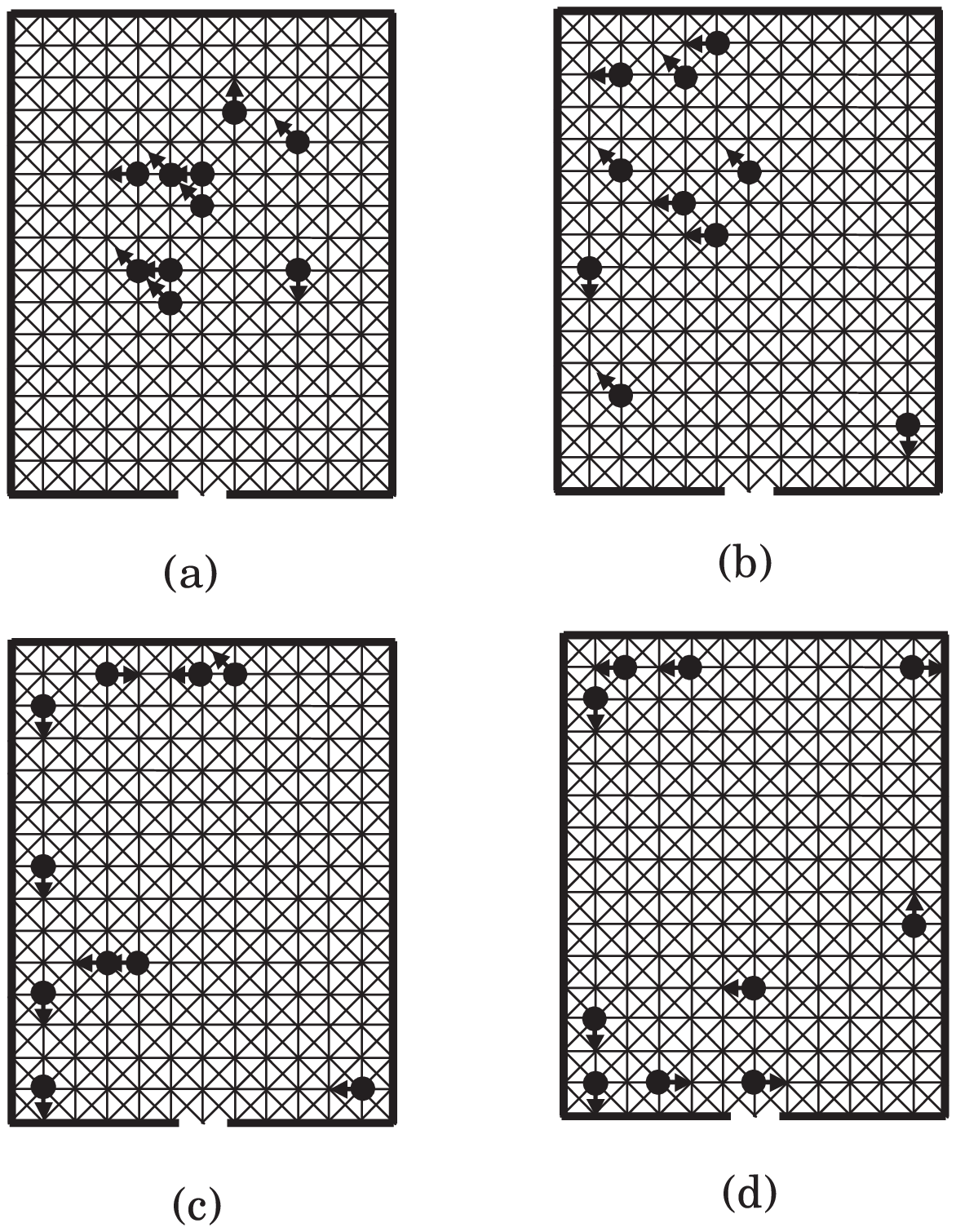}
    \end{center}
\caption[]{\label{Fig9}}
\end{figure}

\clearpage

\begin{figure}[htbp]
    \begin{center}
        \includegraphics[width=12cm]{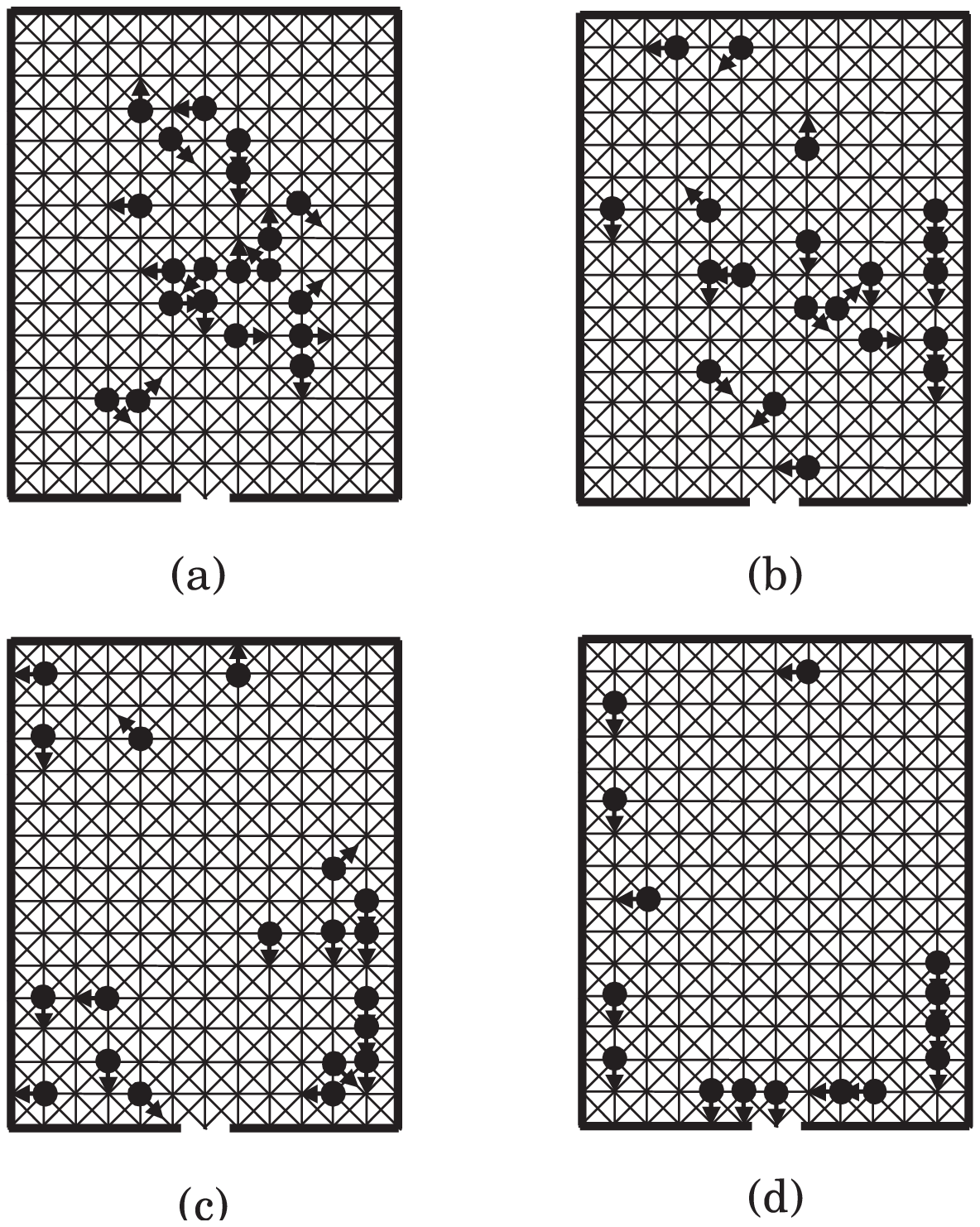}
    \end{center}
\caption[]{\label{Fig10}}
\end{figure}

\begin{figure}[htbp]
    \begin{center}
        \includegraphics[width=12cm]{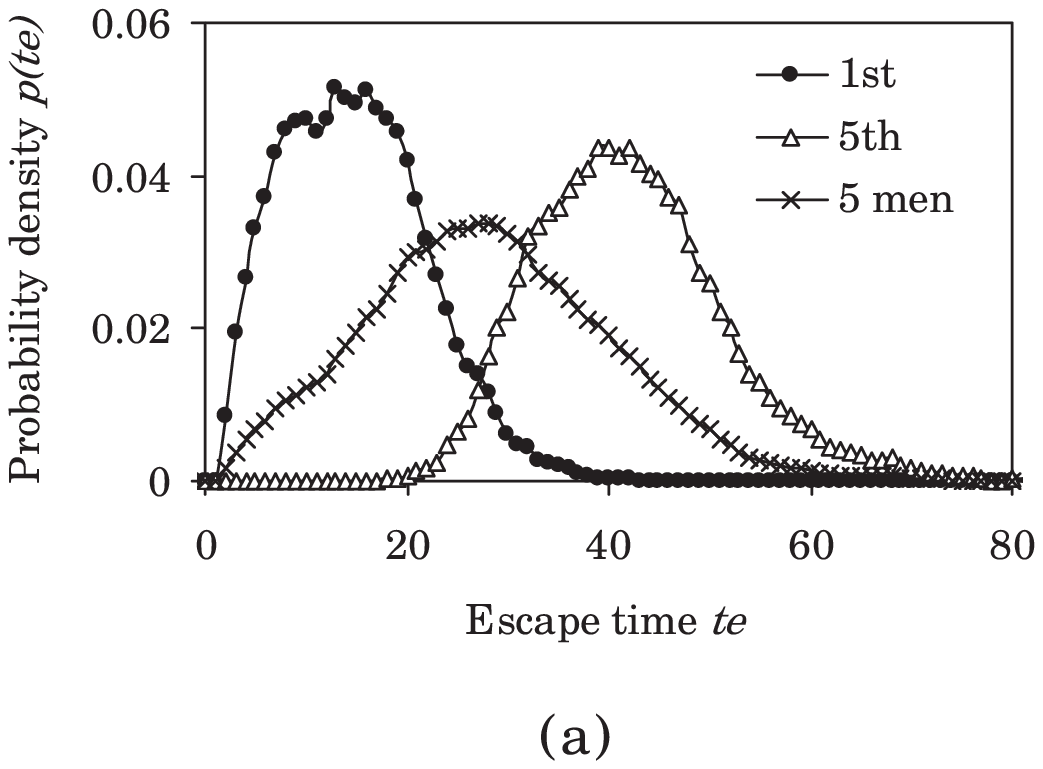}\\
        \includegraphics[width=12cm]{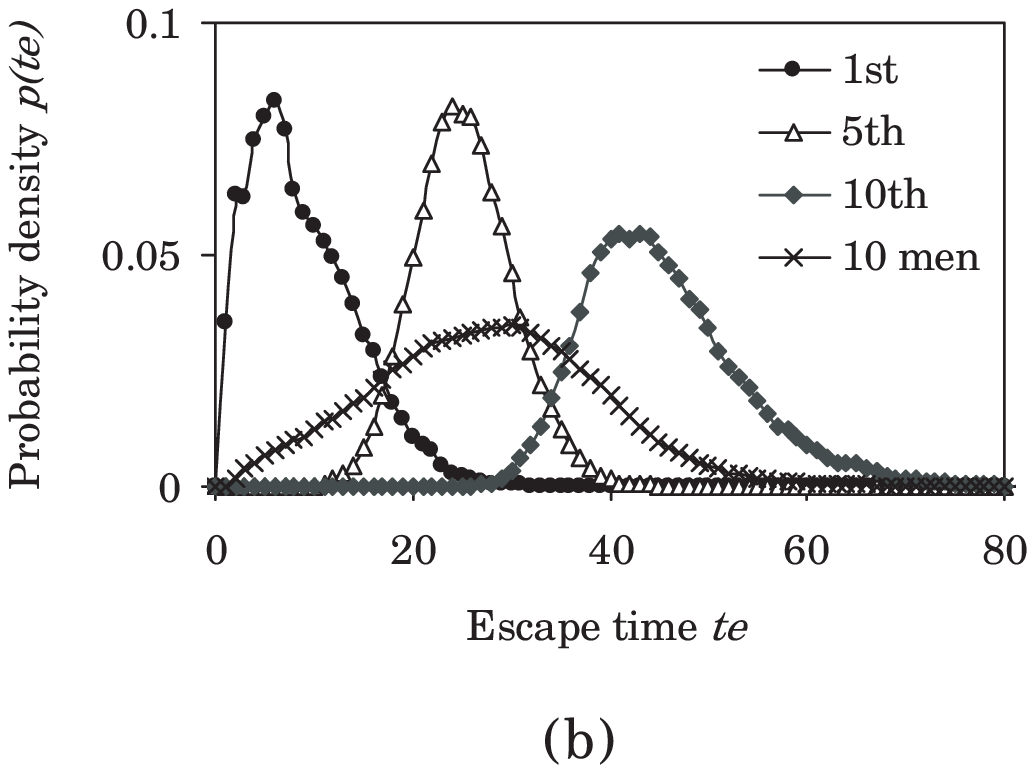}\\
        \includegraphics[width=12cm]{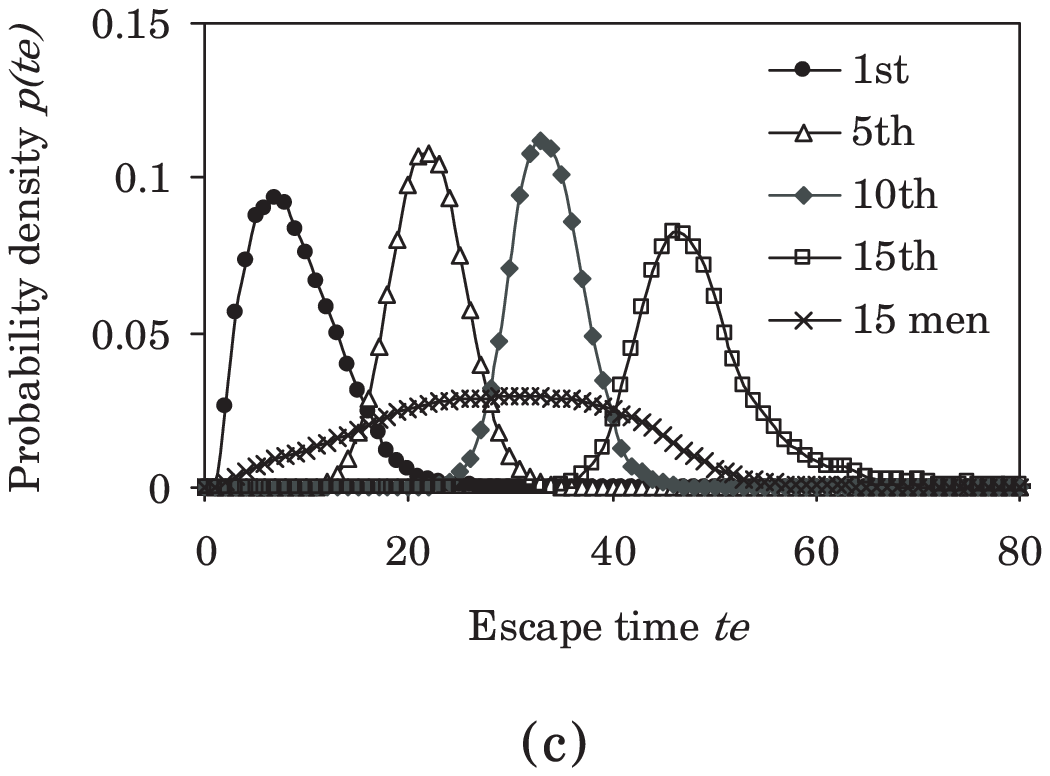}\\
        \includegraphics[width=12cm]{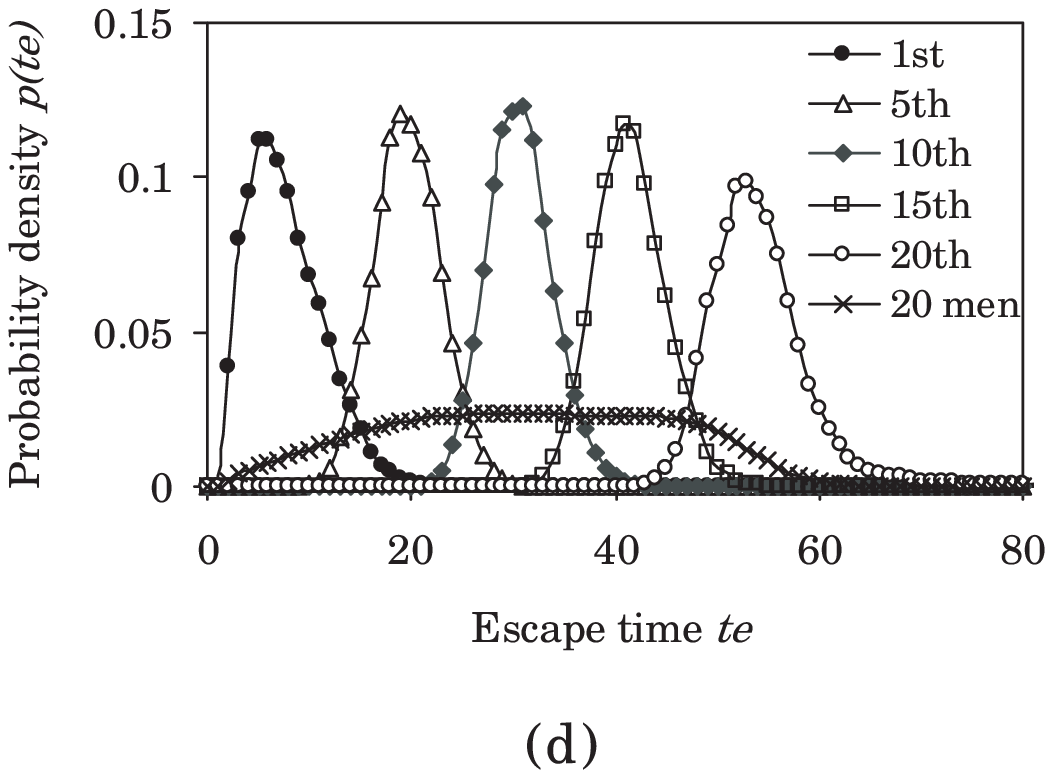}
    \end{center}
\caption[]{\label{Fig11}}
\end{figure}

\clearpage

\begin{figure}[htbp]
    \begin{center}
        \includegraphics[width=12cm]{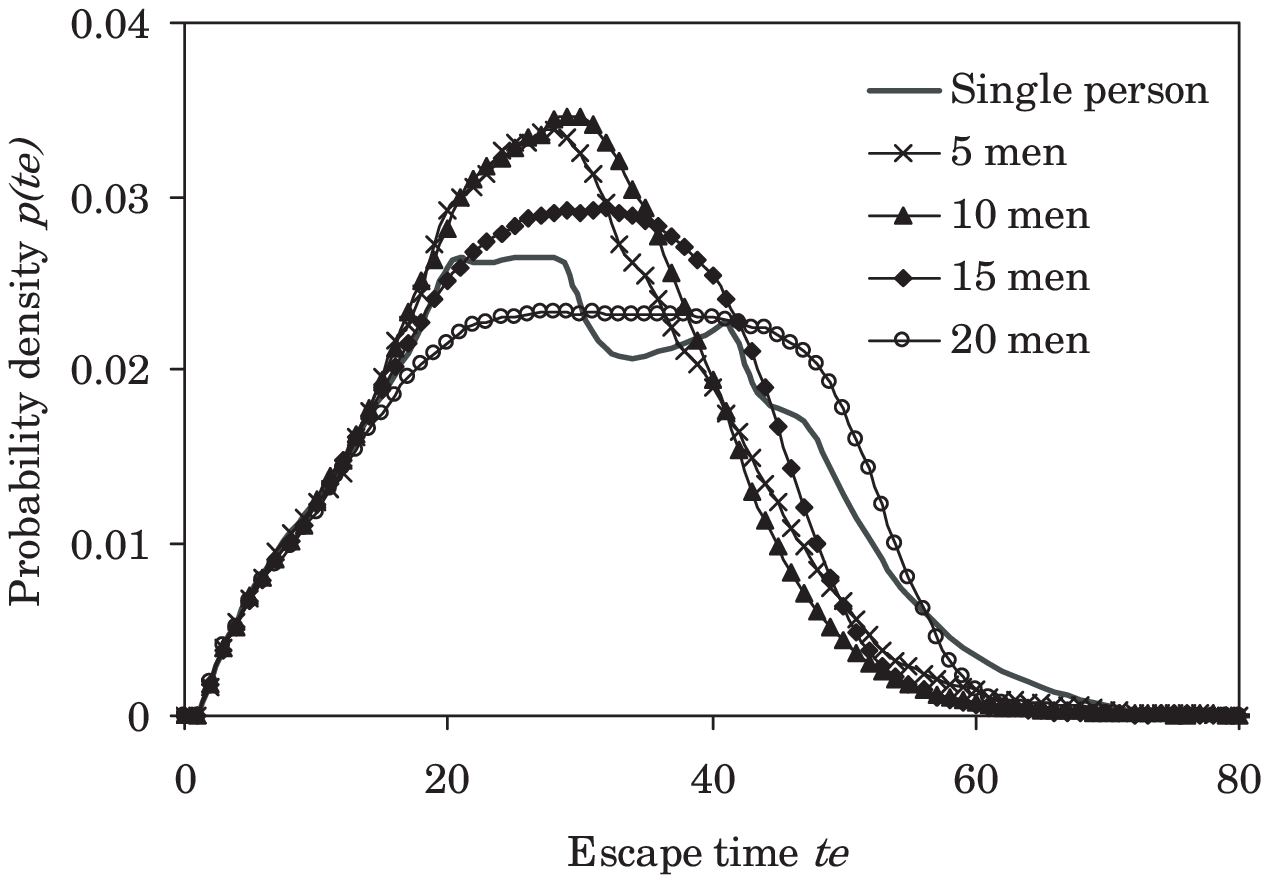}
    \end{center}
\caption[]{\label{Fig12}}
\end{figure}

\begin{figure}[htbp]
    \begin{center}
        \includegraphics[width=12cm]{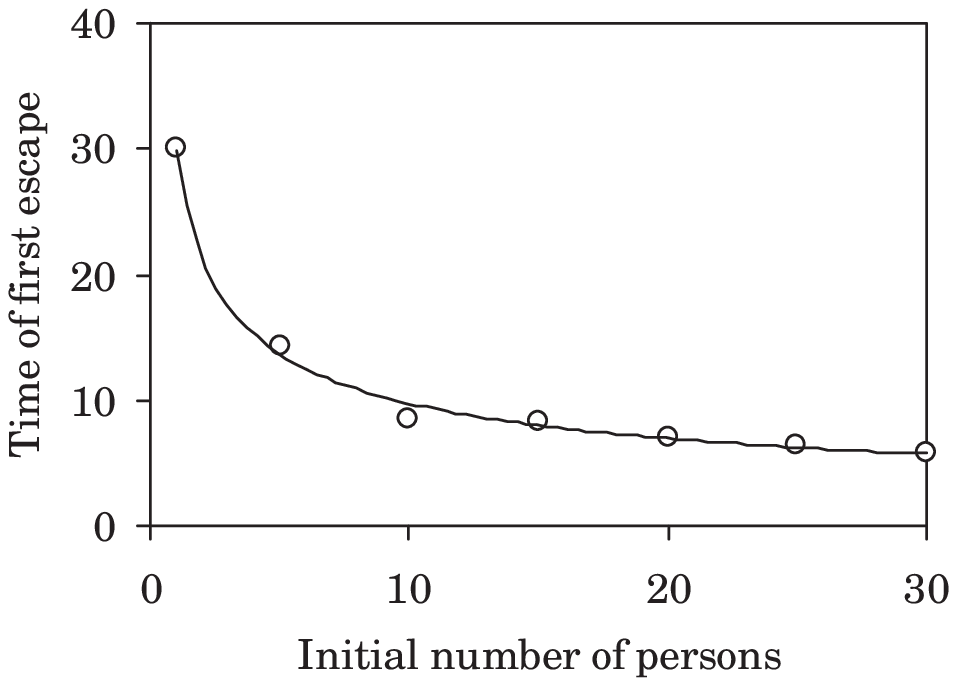}
    \end{center}
\caption[]{\label{Fig13}}
\end{figure}

\begin{figure}[htbp]
    \begin{center}
        \includegraphics[width=12cm]{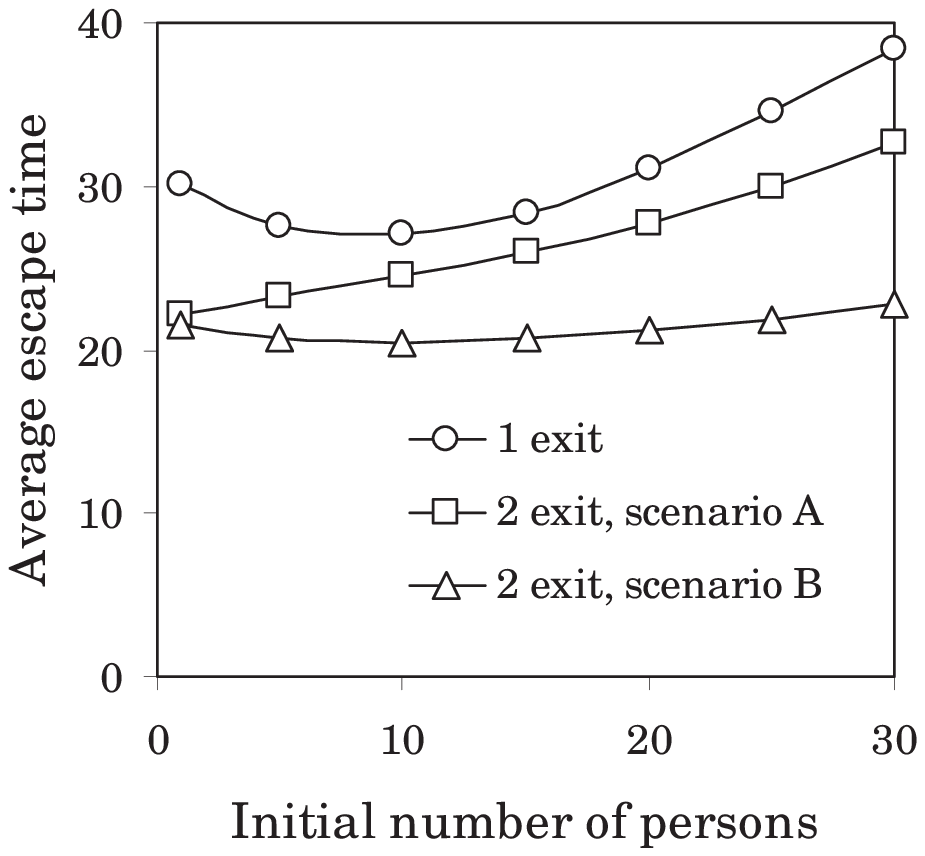}
    \end{center}
\caption[]{\label{Fig14}}
\end{figure}

\begin{figure}[htbp]
    \begin{center}
        \includegraphics[width=12cm]{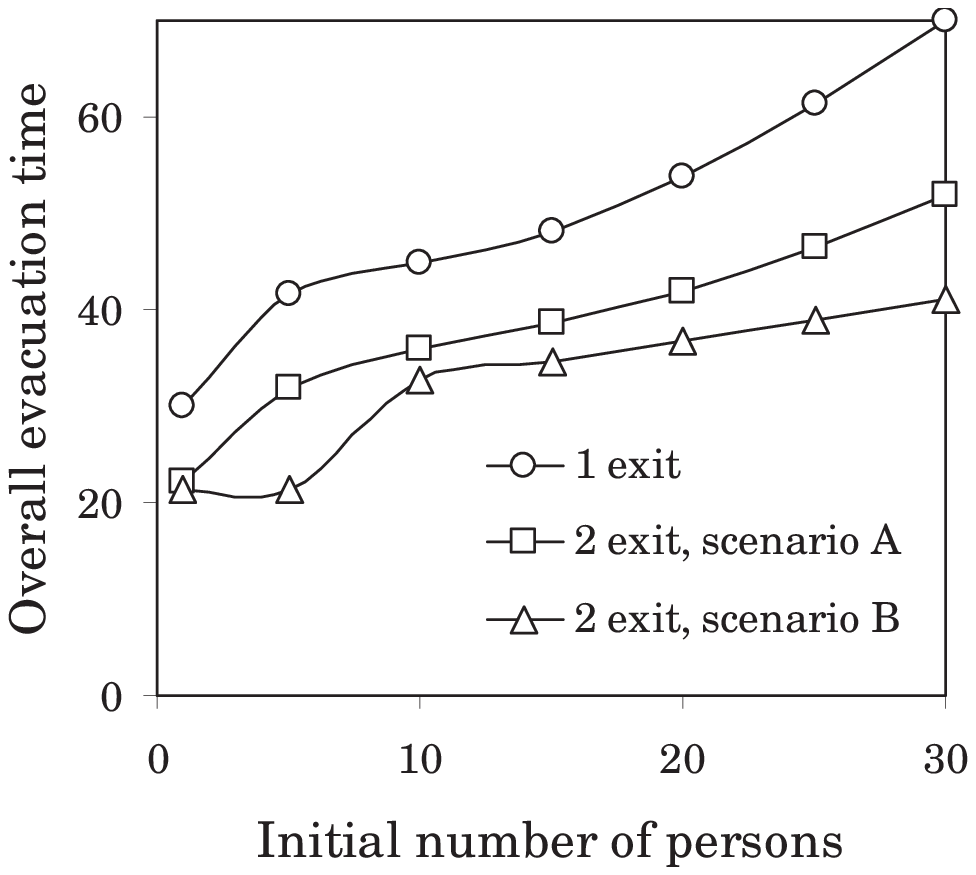}
    \end{center}
\caption[]{\label{Fig15}}
\end{figure}

\end{document}